\documentclass[11pt,letter]{article}
\pdfoutput=1
\usepackage[utf8]{inputenc}

\title{Eventually Consistent Register Revisited}

\include{mathdefs}

\usepackage{titling}
\usepackage{authblk}
\usepackage{mathtools}
\usepackage{amsmath}
\usepackage{amssymb,amsmath,stmaryrd,ifsym}
 \usepackage{graphicx}
\usepackage[protrusion=true,expansion=true]{microtype}
\usepackage[numbers,square,sort&compress]{natbib}
\usepackage{color}
\usepackage{subcaption}
\usepackage{hyperref}  

\newcommand{\rdtOp}[1]{\ensuremath{\mathtt{#1}}}
\newcommand{\oread}{\rdtOp{read}}
\newcommand{\owrite}{\rdtOp{write}}
\auxfun{rd}

\newcommand{\vis}{\mathop{\hbox{$\mathsf{hb}$}}\nolimits}
\newcommand{\arrowvis}{\xrightarrow{\mathsf{hb}}}
\auxfun{op}
\auxfun{rval}
\auxfun{resolve}
\auxfun{deliver}
\newcommand{\spec}{\mathcal{F}}
\newcommand{\type}[1]{\ensuremath{\mathtt{#1}}}
\newcommand{\mvr}{\type{mvr}}
\newcommand{\mvrr}{\type{mvrr}}
\newcommand{\power}{\mathcal{P}}

\begin{document}

\author[1]{Marek Zawirski\thanks{Now at Google.}}
\author[2]{Carlos Baquero}
\author[3]{Annette Bieniusa}
\author[4]{Nuno Preguiça}
\author[1]{Marc Shapiro}

\affil[1]{Inria \& Sorbonne Universités, UPMC Univ Paris 06,
LIP6}
\affil[2]{HASLab, INESC Tec \& Universidade do Minho}
\affil[3]{U.\ of Kaiserslautern}
\affil[4]{NOVA LINCS, DI, FCT, Universidade NOVA de Lisboa}

\date{}
\maketitle

\begin{abstract}
In order to converge in the presence of concurrent updates, modern eventually
consistent replication systems rely on causality information and
operation semantics.
It is relatively easy to use semantics of high-level operations on replicated
data structures, such as sets, lists, etc.
However, it is difficult to exploit semantics of operations on 
\emph{registers}, which store opaque data.
In existing register designs, concurrent writes are resolved either by the
application, or by arbitrating them 
according to their timestamps.
The former is complex and may require user intervention, whereas the latter
causes arbitrary updates to be lost.
In this work, we identify a register construction that generalizes existing ones
by combining runtime causality ordering, to identify concurrent writes, with
static \emph{data semantics}, to resolve them.
We propose a simple conflict resolution template based on an
application-predefined order on the domain of values.
It eliminates or reduces the number of conflicts that need to be resolved by the
user or by an explicit application logic.
We illustrate some variants of our approach with use cases, and how it
generalizes existing designs.
\end{abstract}

\section{Background}
An eventually-consistent replication system accepts updates concurrently at
different replicas.
The challenge is to ensure convergence of values at all replicas under
absence of a common execution order of updates.
To this end, replicas need to interpret delivered updates into a value without
relying on execution order.
Formally, the intended value of an object can be specified in this manner as a
function on the set of delivered updates partially ordered by causality
\cite{crdts-popl}.
Value of abstract data types, such as set, list or counter, can be easily
expressed in this way with the help of their method semantics or causality
relation \cite{crdts-sss}.
This is harder for a low-level register data type with $\owrite$ and
$\oread$ operations, which provide little semantics to make use of.

A classical approach is the multi-value register that uses causality information
to provide all concurrent writes to the application \cite{crdts-sss,vv}.
For the multi-value register that stores values from a domain $V$, the
register value is specified by a function $\spec_\mvr$ that produces a subset
of values from $V$:
\begin{align*}
\spec_\mvr(E, \vis)  = \{ v \mid & \exists e \in E: e =
\owrite(v)
\\
& \land \not \exists e' \in E: e' = \owrite(\_) \land e \arrowvis e'
\},\nonumber
\end{align*}
where $E$ is a set of events observed by read operation,
and $\vis$ is a causality partial order on $E$.
Provided all replicas eventually observe the same set of updates, and
always observe restriction of a common causality relation, the register
converges \cite{crdts-popl}.

When more than one value appears in the set returned by the multi-value
register, it indicates concurrent updates, called a \emph{conflict}.
Conflicts are undesirable, since either the application or the user need
to resolve them, which is complex and may in turn cause another conflict.

\section{Register with Data-Driven Conflict Resolution}
We propose a simple template for conflict resolution based on a predefined order
of values.
This approach reduces or even eliminates the number of conflicts that
need to be resolved by an explicit logic or by the user. 

We define a generalization of the classical multi-value register as
$\spec_\mvrr$:
\begin{align*}
  \spec_\mvrr(E, \vis)  = \resolve(\spec_\mvr(E, \vis)),
\end{align*}
where $\resolve: \power(V) \longrightarrow \power(V)$ is a function that
can resolve some or all of the conflicts.
Hereafter, we identify some simple yet useful classes of $\resolve$.
 

\subsection{Partially Ordered Values}

Let $\prec$ be a strict partial order predefined on values $V$ by the
application, embedded in the object type.
We define $\resolve_\prec$ based on this order as:
\begin{align*}
\resolve_\prec(V)  = \{ v \in V \mid \not \exists v' \in V: v
\prec v' \}.
\end{align*}
The register eliminates concurrently written values that are dominated in
$\prec$. The result is the set of maximal values acording to the order on values.
This reduces the number of conflicts that the user, or the application,
need to resolve. 
When the order $\prec$ is not provided (empty), $\resolve_\prec$ behaves as the
identity function, as in the classical multi-value register.

\subsection{Totally Ordered Values}
A special case of partial order is a total order.
Under total order, $\resolve_\prec$ ensures that the register presents at
most one value to the application, i.e., $\forall X : \left|\resolve(X)\right|
\le 1$.
This is a desirable property, since applications and users are often 
expecting to read a single value, as in the sequential register.

\section{Use Cases}
Instantiations of our construction can be applied to a number of use cases.

\subsection{Semantics-based Ordering}
An application can define the order according to the semantics of stored values.

For example, consider a software bug tracking system.
A register may store priority level of a bug, from a predefined and
totally-ordered domain of priority levels.
Our construction provides a reasonable convergent behavior: concurrent
assignments of different levels should converge to the highest one.
Nevertheless, it allows to decrease the level again, with a later assignment.

A bug tracker may use another register to store status of a bug.
Consider the following status options: \emph{open}, \emph{assigned},
\emph{closed-fixed}, and \emph{closed-irreproducible}.
In this case, the application can specify a partial order on statuses, e.g.,
\emph{assigned} dominates \emph{open}, dominated in turn by the two incomparable
variants of \emph{closed}.
Using this order with our construction, concurrent modifications of the status
converge to a single value, except when the bug is both marked as irreproducible
and fixed, which requires user intervention.

\subsection{Runtime-based Ordering}
Although the order of values $\prec$ is static, it can be also based on
runtime-provided information, such as replica ID or timestamp.
In particular, our construction can achieve behavior similar to the
last-writer-wins policy (LWW) \cite{lww}, provided every $\owrite(v)$ operation
is augmented with a timestamp $t$ at the time of write, becoming effectively
$\owrite((v, t))$, and pairs $(v, t)$ are totally-ordered according to their
$t$.

An advantage of our approach compared to the classical LWW register is that the
timestamps are used to arbitrate the concurrent values only, avoiding some of
the arbitration anomalies caused by physical clocks  \cite{lww-trobules}.
For instance, it is no longer possible to timestamp a write, with a far future
time, and prevent later writes to appear.
Any write that observes this write will be, in our construction, ordered after
that write, regardless of the timestamp.

\section{Implementation}



We illustrate an implementation of the proposed register in the state-based
eventually-consistent replication model \cite{crdts-sss}.
In this model, replicas opportunistically exchange their complete states via
message passing.

\begin{figure}[h]
{\scriptsize
\begin{tabular}{lrcl}
Replica states & $\Sigma$ & $=$ & $\mathcal{P}(\ids \times \nat \times V) \times (\ids \func \nat)$ \\ 
Initial state & $\sigma_i^0$ & $=$ & $(\{\},\{\})$ \\
Write at replica $i$ & $\owrite_i\big(v,(s,c)\big)$ & $=$ &  $\big( \{(i,c[i]+1,v)\}, c[i
\mapsto c[i]+1]\big)$\\
Read at replica $i$ & $\oread_i\big((s,c)\big)$ & $=$ & $\{v | (\_,\_,v) \in s\}$\\
Merge replica states &$\deliver\big((s,c), (s',c')\big)$ & $=$ &
$\mathsf{resolve}_\prec\big((s \cap s') \cup \{ (i,n,v) \in s | n > c'[i] \} $ 
\\&&&\qquad$ \cup\{ (i',n',v') \in s' | n' > c[i'] \}, c \join c'\big)$ \\
\qquad\quad\qquad where &$\mathsf{resolve}_\prec\big((s,c)\big) $& $=$ & $\big(\{
(i,n,v) \in s | \not\exists (\_,\_,v') \in s \cdot v \prec v'\},c\big)$\\
\end{tabular} } \caption{Optimized implementation of register with
$\mathsf{resolve}_\prec$, replica $i$.}
\label{fig:mvregpo}
\end{figure}

The register implementation in Figure~\ref{fig:mvregpo} uses an implementation
of $\mathsf{resolve}_\prec$ to reduce any concurrently assigned values
according to the partial order $\prec$ defined by the application on those
values.
The order among values can range from:
No ordering \--- all values are concurrent, and thus not order reducible;
Partial order \--- one or more maximal values are kept after resolve; Total
order \--- a single maximal value is kept after resolve.
The algorithm includes an optimization that allows storing a single scalar
logical clock to identify each written value, complemented by a version vector
for the whole register.
The classical multi-value register implementation stores a version vector per
value \cite{vv,crdts-sss}.

The state is composed by a set of values, tagged by scalar clocks, and by a
common version vector.
The scalar clocks are locally generated by using a replica id $i \in \ids$ and a
monotonic counter per replica.
A write operation $\owrite_i(v,\sigma)$ is depicted as a state transforming
function, tagged with the replica id $i$, and supplying a value $v$ and the
current state $\sigma=(s,c)$, where $s$ is the set and $c$ is the ``causal
context'' version vector.
Each write uses the version vector to create a new scalar clock and derives a
new set with a single value tagged by the scalar clock, as well as an updated
version vector that includes the new scalar.
The read operation $\oread_i(\sigma)$ keeps the state unchanged and replies with
a set comprising all values present in the multi-value register, stripped of
clock metadata.

Since writes always derive a set with a single value, the set will only have
multiple values as a result of a merge that gathers concurrently assigned
values, written in different replica states.
The merge collects concurrently assigned values that have not been
overwritten and supplies these values to the $\mathsf{resolve}_\prec$ for
possible further reduction on resulting set.
The implementation in $\deliver$ detects values that have been observed and
later overwritten by checking that the scalar cloks associated to those values are
included in the version vector $c$ while those entries are no longer present in
the set $s$.
Values still present on both sets, or newly written values are kept.
This detects and keeps all concurrently assigned values, but when
$\mathsf{resolve}_\prec$ is finally called some of these values can be removed
if the order information on values indicates that they are dominated by a higher
value.

\begin{figure}[h]
\centering
\includegraphics[scale=0.7]{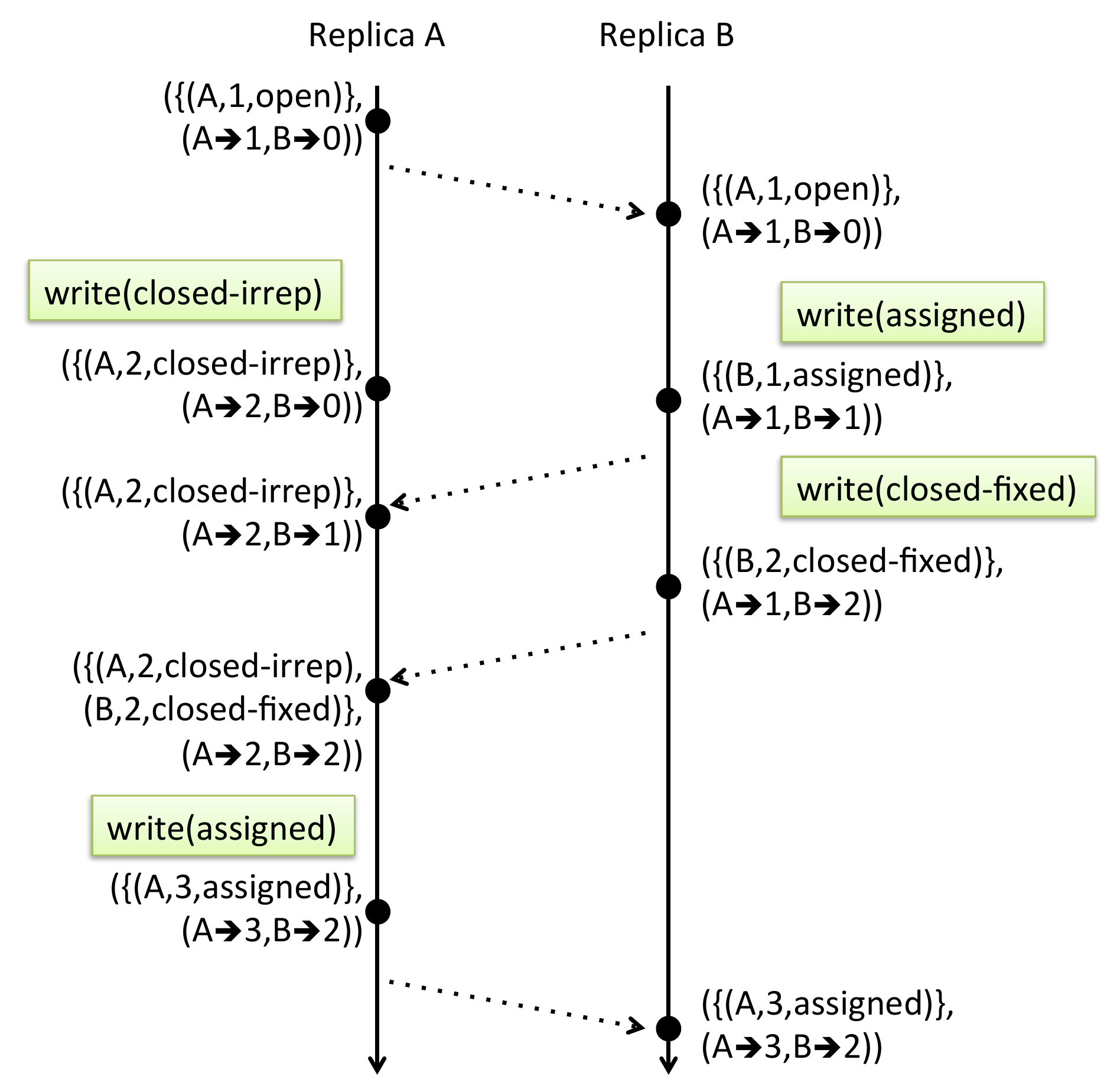}
\caption{Bug tracking run with two register replicas; dashed arrow represents 
a message, merged at the receiver replica; solid box indicates a client
operation.}
\label{fig:rdt-bug}
\end{figure}

Figure~\ref{fig:rdt-bug} shows a run of a system with two replicas for the 
bug tracking example mentioned before. After the first synchronization from
replica B to replica A, the state will be \emph{closed-irrep}, as this value is
greater than \emph{assigned} in the order of values. After the second synchronization,
the register will maintain two values as \emph{closed-irrep} and 
\emph{closed-fixed} are incomparable.
Later, these values are replaced by a new write with value \emph{assigned}.

\begin{figure}[h]
\centering
\includegraphics[scale=0.7]{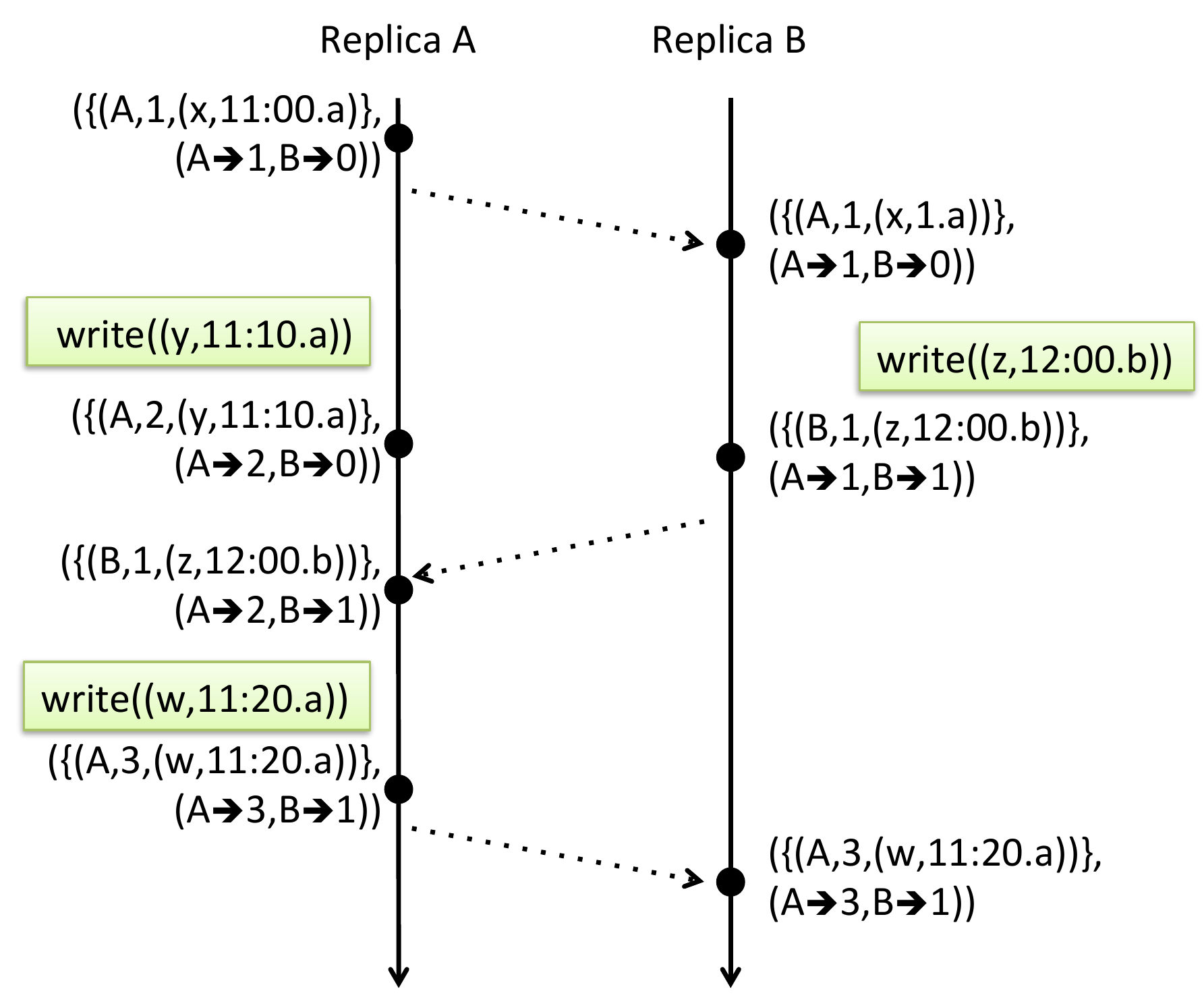}
\caption{Last-writer-wins run with two register replicas; dashed arrow represents 
a message, merged at the receiver replica; solid box indicates a client
operation.}
\label{fig:rdt-lww}
\end{figure}

Figure~\ref{fig:rdt-lww} shows a run with the last-writer-wins behavior.
In this example, we assume that replica B has a local clock at a higher
value.
We can see that after the first write in replica B is propagated to replica A,
the following write in A will overwrite the value previously written by replica
B, although the new timestamp is smaller. The reason for this is that the 
timestamp is only used to arbitrate among concurrent values.

\paragraph{Acknowledgments} 
This work was partially supported by EU FP7 SyncFree project (609551),
FCT/MCT projects PEST-OE/EEI/UI0527/2014 and UID/EEA/50014/2013, and
by a Google Faculty Research Award 2013.

\newpage
\bibliographystyle{abbrvnat}
\bibliography{short-predef,local,shapiro-bib,bib}

\end{document}